# Luminescence properties of point defects in silica

Marco Cannas

*INFM, Dipartimento di Scienze Fisiche ed Astronomiche, via Archirafi 36 I-90123 Palermo*
*Phone: +39-091-6234220; Fax: +39-091-6162461; E-mail: cannas@fisica.unipa.it*

**Abstract.** The optical properties of point defects in as-grown natural silica are reviewed. Two emissions peaked at 4.2 eV ($\alpha_E$ band) and at 3.1 eV ($\beta$ band), related to an absorption band at 5.1 eV ($B_{2\beta}$), have been experimentally investigated on the basis of their temperature dependence and their kinetic decay. Our results allow to characterize the excitation pathway of these luminescence bands and to make clear the competition between the radiative relaxation rates and the phonon assisted intersystem crossing process linking the singlet and the triplet excited states from which $\alpha_E$ and $\beta$, respectively, originate. Finally, we discuss the role played by the disordered vitreous matrix in influencing the optical features of defects.

## 1. Introduction

The study of vitreous silica, the amorphous silicon dioxide ($a$-SiO$_2$), is currently an attractive research field in solid state physics and material science [1-3]. The physical properties of high transparency in a wide spectral region (visible, UV, vacuum-UV) and low conductivity, in combination with favorable mechanical characteristics and low manufacturing costs, have led to the widespread utilization of silica-based materials in many technological applications, like manufacturing of optical fibers, lenses and optoelectronic devices.

These exceptional features depend critically on the maintenance of defect free band gap [4]. For this reason the understanding of the nature and the formation mechanisms of defects in $a$-SiO$_2$ plays a fundamental role in both the technological and basic research. To this end, the combined use of several spectroscopic techniques and silica materials different for their manufacturing processes or for external treatments (irradiation, heating) could provide a powerful method to improve the knowledge of the properties of defects. In this paper we focus our attention on the optical absorption (OA) and photoluminescence (PL) bands detected in as grown natural silica in a wide spectral region from visible to UV. Aim of this work is to exemplify the whole optical activity in terms of an energy level diagram with the radiative and non radiative transitions accounting for the spectral and kinetics features, also as a function of temperature. Moreover, we investigate the role played by the vitreous matrix in determining the observed optical activities. It is worth noting that the study of defects in silica holds a more general significance in obtaining new insights on the relationship between the properties of an optically active defect and the structural and dynamic properties of its environment in other disordered materials.

The present paper is so structured: in the section 2 we outline the theoretical background on the optical properties of a generic point defect in silica, in section 3 we describe the experimental method and finally in the section 4 we review and discuss our experimental results.

## 2. Theoretical background

### *2.1 Point defect in silica*

The point defect is usually defined in the contest of a crystalline network, if the lattice site is occupied differently than in the perfect crystal [5]. The defects may be classified as intrinsic and extrinsic. The first type includes unoccupied sites (vacancies) and occupied sites that in the perfect crystal are unoccupied (interstitial). The second type consists of impurities at sites that in the crystal lattice either are occupied by atoms of the pure material (substitutional impurities) or are unoccupied (interstitial impurities). An overview of the different kinds of defects is shown in Fig. 1.

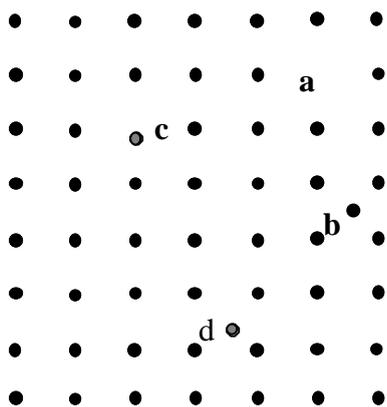

**Fig. 1** Crystal lattice with different types of defects: intrinsic vacancy (a) and interstitial (b); extrinsic substitutional (c) and interstitial (d).

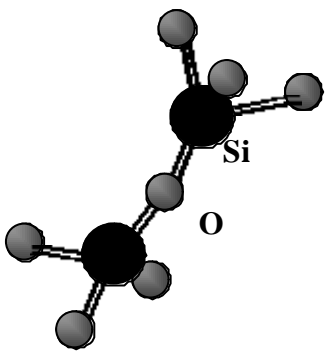

**Fig. 2** Fragment of a regular silica network

The concept of defect can also be extended to amorphous materials, like silica, whose structure matches the crystalline α-quartz only in a short-range order [2]. As depicted in Fig. 2, the structural unit of silica is the $SiO_4$ tetrahedron where the Si atom is bonded to four O atoms with O-Si-O angle of 109.5°. The lack of long-range atomic order is due to the large spread of tetrahedral linkage angle Si-O-Si, statistically distributed between 120° and 180°. In this framework, a point defect is present when the array of Si and O atoms of the ideal silica network is broken down by an imperfection.

The presence of defects in the vitreous matrix may drastically modify the optical properties of the host material [4, 6, 7]. Indeed, defects exist in different electronic states that can cause optical transitions as absorption and luminescence with lower energies than the fundamental absorption edge of the silica material, approximately 9 eV, from valence to conduction band. For this reason, a point defect is also defined a color center or chromophore.

Even if these transitions are localized at the point defect, the optical spectra are influenced by its environment. While in an isolated center, the energy transferred in an optical transition has to match the difference between the electronic levels whose spread is limited only by the excited state lifetime, for a chromophore embedded in a matrix, it can be shared between a local electronic contribution and a wide variety of phonon excitations of the vibrational modes of the matrix. Moreover, owing to the amorphous nature, each defect can exist in different local rearrangements of the surrounding matrix (conformational inhomogeneity) and the energy associated to transitions between the electronic levels is largely distributed. Then, a wide range of photon energies can be involved in the transition and the spectrum of absorption or luminescence consists of broad bands.

## 2.1 Optical absorption transition

We consider now the transition occurring between two electronic states of a point defect: the ground state (0) and the excited one (1) [5, 8, 9]. In Fig. 3, we sketch the potential energy curves $\varepsilon_0$ and $\varepsilon_1$ as a function of the generalized normal coordinate $Q_f$. In the same figure are also depicted the vibrational levels associated to the quantum numbers $n_f$ and $m_f$ in the states (0) and (1), respectively, due to the nuclear oscillations. If the equilibrium positions of atoms in the ground and excited state are different, the set of normal coordinates changes from $Q_f$ (in the ground state) to $Q'_f = Q_f - D_f$ (in the excited state). Then, the total energies associated to the (0) and (1) states can be expressed by:

$$E(0)_{Tot} = \varepsilon_0^{eq} + \sum_{f=1}^{N_f} h\nu_f \left(n_f + 1/2\right) \qquad n_f = 0, 1, 2, \ldots \qquad (1)$$

$$E(1)_{Tot} = \varepsilon_1^{eq} + \sum_{f=1}^{N_f} h\nu_f \left(m_f + 1/2\right) \qquad m_f = 0, 1, 2, \ldots \qquad (2)$$

where $\varepsilon_0^{eq}$ and $\varepsilon_1^{eq}$ are the energies of the two electronic states when all the nuclei are in their equilibrium position. In the above Eqs., we have assumed that the frequency $\nu_f$ of normal modes is the same in the two states (0) and (1) (linear coupling approximation).

According to the *Franck-Condon* approximation, that is, the electronic transitions occurring in a time much faster than the nuclear motion, the absorption is described as a vertical transition with respect to the energies associated to the two states in the configuration coordinates diagram. If $E_{1,0} = E_1 - E_0 = h\nu_{1,0}$ is the energy value matching the quantum transition between the two states (0) and (1), for a $dl$ path length of a sample having $N_0$ identical non-interacting absorbers per unit volume in the ground state, the differential energy loss by the electromagnetic field is given by [8]:

$$-dI(E) = I(E) N_0 \frac{4\pi^2}{3\hbar^2} \frac{1}{4\pi\varepsilon_0} \frac{E}{c} \cdot \delta(E - E_{1,0}) |R_{1,0}|^2 dl \qquad (3)$$

where $I(E)$ is the intensity of the electromagnetic field flowing through the sample and $R_{1,0}$ is the quantum-mechanical matrix element of the electric dipole moment **M** between the total eigenfunctions $\psi_0$ and $\psi_1$ of the two states (transition moment).

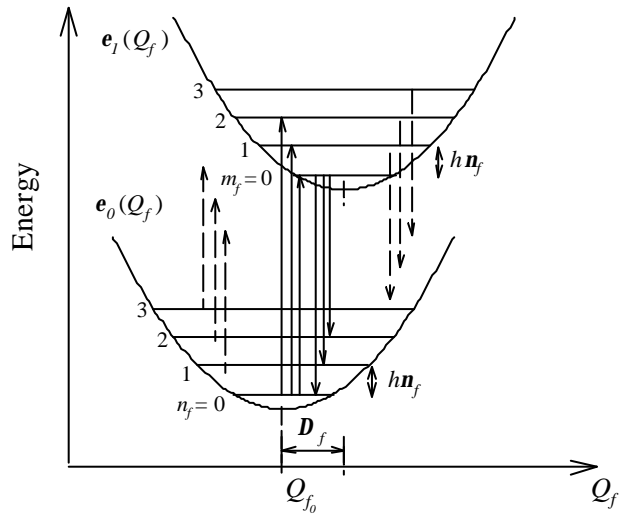

**Fig. 3** Optical absorption and luminescence transitions between the ground ($\varepsilon_0$) and excited ($\varepsilon_1$) electronic states in a configuration coordinate diagram

$$|R_{1,0}|^2 = |\langle \psi_1|\mathbf{M}|\psi_0\rangle|^2 \quad (4)$$

By integration over a unitary path of absorbing material, one obtains the following expression for the absorption coefficient:

$$\alpha(E) = Ln\frac{I_0(E)}{I(E)} = \frac{N_0 \pi E}{3\hbar^2 \varepsilon_0 c}|R_{1,0}|^2 \delta(E - E_{1,0}) \quad (5)$$

where $I_0(E)$ is the intensity of the electromagnetic field incident on the sample at energy $E$. Taking into account the finite lifetime $\tau$ of the excited state, the $\delta$ function can be replaced by a *Lorentzian* shape function (with $2\Gamma = 1/\tau$):

$$\alpha(E) = \frac{N_0 \pi E}{3\hbar^2 \varepsilon_0 c}|R_{1,0}|^2 \frac{\Gamma}{(E - E_{1,0})^2 + \Gamma^2} \quad (6)$$

In order to obtain a suitable expression of the transition moment $R_{1,0}$, we make use of the *Born-Oppenheimer* approximation and write the eigenfunction $\psi$ of the total system by the product of the electronic wavefunction $\phi$ and the nuclear one $\chi$ [5, 8]:

$$\psi_0(r,Q) = \phi_0(r,Q)\cdot\chi_0(Q) \quad (7)$$
$$\psi_1(r,Q') = \phi_1(r,Q')\cdot\chi_1(Q') \quad (8)$$

where $r$ is the electronic coordinate and $Q$ and $Q'$ are the set of normal nuclear coordinates in (0) and (1), respectively, which allow to write $\chi_0$ and $\chi_1$ as a product of harmonic oscillator functions:

$$\chi_0(Q) = \prod_{f=1}^{N_f}\chi(Q_f); \quad \chi_1(Q') = \prod_{f=1}^{N_f}\chi(Q_f - \Delta_f) \quad (9)$$

We can consider the electronic dipole moment **M** as composed by an electronic term, $\mathbf{M}_{el.}(r)$, and nuclear one, $\mathbf{M}_{nucl.}(Q)$ [8]. So, we can write:

$$R_{1,0} = \iint \psi_1^*(r,Q')\mathbf{M}\psi_0(r,Q)d\tau_e d\tau_v =$$
$$\iint \phi_1^*(r,Q')\chi_1^*(Q')\mathbf{M}_{el.}(r)\phi_0(r,Q)\chi_0(Q)d\tau_e d\tau_v +$$
$$+ \iint \phi_1^*(r,Q')\chi_1^*(Q')\mathbf{M}_{nucl.}(Q)\phi_0(r,Q)\chi_0(Q)d\tau_e d\tau_v \quad (10)$$

where the integral is carried out on the space of the electronic $\tau_e$ and vibronic $\tau_n$ coordinates. Since during the transition from (0) to (1) the nuclei remain almost stationary in the equilibrium position of the ground state, the electronic eigenfunction $\phi$ can be assumed to depend upon $Q=Q_0$. Besides, we note that $\mathbf{M}_{nucl.}(Q)$ does not depend on the electronic coordinates but only upon vibrational coordinates $Q$. In this way, the expression for $R_{1,0}$ can be readjusted as follows:

$$R_{1,0} = \int \phi_1^*(r,Q_0)\mathbf{M}_{el.}(r)\phi_0(r,Q_0)d\tau_e \cdot \int \chi_1^*(Q')\chi_0(Q)d\tau_v +$$
$$+ \int \phi_1^*(r,Q_0)\phi_0(r,Q_0)d\tau_e \cdot \int \chi_1^*(Q')\mathbf{M}_{nucl.}(Q)\chi_0(Q)d\tau_v \quad (11)$$

Owing to the orthogonality of electronic eigenfunctions, the second term vanishes and the transition moment $R_{1,0}$ is given by:

$$R_{1,0} = R_e \cdot \int \chi_1^*(Q')\chi_0(Q)d\tau_v \quad (12)$$

where $R_e$ indicates the transition moment associated to the electronic states $\phi_0$ and $\phi_1$:

$$R_e = \int \phi_1^*(r,Q_0)\mathbf{M}_{el.}(r)\phi_0(r,Q_0)d\tau_e \quad (13)$$

The term $\int j_1^*(Q')j_0(Q)dt_v$, known as the *Franck-Condon* integral, measures the overlap between the vibrational functions $j_0$ and $j_1$. If there is no coupling between the electronic transition and the vibrational modes, the set $Q'$ coincides with $Q$ ($D_f = 0$) and, owing to the orthogonality of harmonic oscillator wavefunctions, only terms with the same vibrational quantum numbers (i.e. $m_f = n_f$) contribute to the transition. In the presence of coupling, $Q' \neq Q$ and also terms with $m_f \neq n_f$ will contribute to the transition.

Taking into account the Eqs. (9) and (12), the absorption coefficient is given by:

$$a(E) = M_0 \, E \left| \prod_{f=1}^{N_f} \int j_f^*(Q - \Delta_f) j_f(Q) dt_v \right|^2 \frac{\Gamma}{(E - E_{1,0})^2 + \Gamma^2} \quad (14)$$

with $M_0 = \dfrac{N_0 p}{3\hbar^2 e_0 c} |R_{1,0}|^2$.

At T=0 K, only the vibrational level with $n_f=0$ in the ground electronic state (0) is populated and Eq. (14) can be rewritten as:

$$a(E, T=0) = M_0 \, E \cdot \sum_{\{m_f\}} \left[ \left( \prod_{f=1}^{N_f} e^{-S_f} \frac{S_f^{m_f}}{m_f!} \right) \times \frac{\Gamma}{\left(E - E_{00} - h\sum_{f=1}^{N_f} m_f n_f\right)^2 + \Gamma^2} \right] \quad (15)$$

where $E_{1,0}$ has been expressed in agreement with Eqs. (1) and (2) to take into account the energies associated to the transitions from $n_f = 0$ in the state (0) to various $m_f$ in the state (1), as shown by the upwards arrows (continuous line) in Fig. 3. In particular, $E_{00}$ represents the energy difference between the $m_f = 0$ level in (1) and $n_f = 0$ in (0), i.e. the energy of the purely electronic transition. The dimensionless *Huang-Rhys* factor $S_f$ [10] (linear coupling constant) is defined by:

$$S_f = \frac{h n_f}{2} \Delta_f^2 \quad (16)$$

and it measures the coupling strength between the electronic transition 0→1 and the *f*–th vibrational mode that determines the rearrangement of the nuclei from their equilibrium position by $D_f$. Then, Eq. (15) shows that the absorption band at T=0 K results from the superposition of a series of *Lorentzians* whose intensity is modulated by the *Poissonian* distributions product:

$$\prod_{f=1}^{N_f} e^{-S_f} \frac{S_f^{m_f}}{m_f!} \quad (17)$$

At T ≠ 0, the vibrational levels with $n_f \neq 0$ in the ground state (0) can be thermally populated according to the *Boltzmann* law and contribute to the transitions toward the excited state (1), as shown by the dashed line arrows in Fig. 3. Generally, if $T_{max.}$ is the upper limit of the temperature range investigated, it is possible to distinguish between vibrational modes with high ($n_h$) and low ($n_l$) frequency [11]. The $n_h$ modes have frequency such that $h n_h \gg K_B T_{max}$, where $K_B$ is the *Boltzmann* constant, so that only the vibrational level with $n_h = 0$ is occupied in the temperature range up to $T_{max}$. This fact implies that, assuming a set of $N_h$ high frequency modes, only transitions from $n_h = 0$ in the state (0) to $m_h = 0, 1, 2,\ldots$ in the state (1) occur. At variance, the $n_l$ modes can change their population on varying the temperature. Therefore, assuming a set of $N_l$ low frequency modes, transitions from $n_l = 0, 1, 2,\ldots$ in (0) to $m_l = 0, 1, 2,\ldots$ in (1) take place and are relevant in changing the shape of the absorption spectrum. If we consider the contribution of these $N_l$ modes as a single mode (*Einstein* oscillator) with mean frequency value $\langle n_l \rangle$ and mean linear coupling constant $S_l$, the expression of the absorption at the temperature T is given by:

$$a(E,T) = M_0 \, E \left[ \sum_{\{m_f\}} \left( \prod_{f=1}^{N_h} e^{-S_f} \frac{S_f^{m_f}}{m_f!} \right) \right] \times$$

$$\times \frac{\Gamma}{\left( E - E_{00} - h \sum_{f=1}^{N_f} m_f \boldsymbol{n}_f \right)^2 + \Gamma^2} \otimes \frac{1}{W(T)} \cdot e^{-\frac{1}{2} \cdot \frac{E^2}{W^2(T)}} \quad (18)$$

where $\otimes$ represents the convolution operator, $f(E) \otimes g(E) = \int f(E-E') \cdot g(E') dE'$, and where

$$W^2(T) = N_l S_l h^2 \langle \boldsymbol{n}_l \rangle^2 \coth \frac{h \langle \boldsymbol{n}_l \rangle}{2 K_B T} \quad (19)$$

According to Eqs. (17), (18) and (19), the absorption profile is given by the superposition of a series of *Voigtians* (*Gaussian* convolutions of *Lorentzians*) whose width increases as the temperature increases while the energy peak remains constant. We recall that the Eq. (18) for the absorption spectrum has been obtained in the linear coupling approximation, i.e. assuming the same vibrational frequency $\boldsymbol{n}_f$ of normal modes in the states (0) and (1). At variance, if the transition from (0) to (1) changes the vibrational frequencies $\boldsymbol{n}_f$ (non linear coupling), temperature effects on the peak position of the absorption band are also present and must be taken into account [11, 12].

Further contributions to the absorption lineshape are the inhomogeneous effects arising from the different local environments surrounding the point defects in amorphous materials. Generally, this site-to-site non-equivalence results in a spectral distribution of the purely electronic transition energies $E_{00}$. If the mapping between the conformational and the spectral heterogeneity is linear [11], the distribution function for $E_{00}$ is given by a *Gaussian* function:

$$g_{abs}(E_{00}) = \frac{1}{\sqrt{2 \boldsymbol{p}} \boldsymbol{s}_{inh}} e^{-\frac{(E_{00} - E'_{00})^2}{2 \cdot \boldsymbol{s}_{inh}^2}} \quad (20)$$

peaked at the mean energy $E'_{00}$ whose width $\boldsymbol{s}_{inh}$ does not depend on the temperature. So, the whole optical absorption spectrum is given by the convolution of Eqs. (18) and (20) and its total width is determined by the different weights of the broadening mechanisms: lifetime, electron-phonon interaction, inhomogeneous broadening.

*2.3    Photoluminescence activity*

Following light absorption, the inverse transition from the excited state (1) to the ground state (0), shown in Fig. 3, causes a spontaneous emission of light also called photoluminescence. In particular, for two states having the same spin multiplicity, the electronic transition is allowed and in this case it is called fluorescence. As in the excited state the nuclei relax towards the minimum energy configuration at $Q_f = Q_{f0} + \Delta_f$ in a much shorter time ($10^{-12}$ s) than the fluorescence lifetime ($10^{-8}$ s), the light emission occurs after that the state (1) has reached the thermal equilibrium [9]. Because the excitation of phonons reduces the energy available to the photon, the luminescence emission occurs at lower energies than the absorption.

The expression of the intensity of light emitted from the excited state (1) is obtained by the relation between the *Einstein* coefficients for absorption and spontaneous emission [13]. If $E_{0,1} = E_1 - E_0 = h \boldsymbol{n}_{0,1}$ is the energy value matching the quantum transition (1)→(0), the luminescence intensity of a sample having $N_{lum}$ identical non-interacting centers per unit volume in the excited state is given by:

$$I_{PL}(E) = N_{lum} \frac{1}{3\hbar^4 \pi \varepsilon_0 c^4} |R_{0,1}|^2 E^4 \frac{\Gamma}{(E-E_{0,1})^2 + \Gamma^2} \quad (21)$$

where $R_{0,1}$ is the quantum-mechanical matrix element of the electric dipole moment **M** for the transition between the states (1) and (0):

$$|R_{0,1}|^2 = |\langle \psi_0 | \mathbf{M} | \psi_1 \rangle|^2 \quad (22)$$

By comparing the Eqs. (21) and (22) with the Eqs. (6) and (4), respectively, it is possible to see that the emission band profile is mirror like to the absorption one apart from a shift of the peak position toward lower energies. Because the excited state (1) is at thermal equilibrium during the luminescence emission, its vibrational levels are populated according to the *Boltzmann* law. At T=0 K, only transitions from the vibrational level with $m_f=0$ in the state (1) to different vibrational levels with various $n_f$ of the state (0) can occur, as depicted by downwards arrows (continuous line) in Fig. 3. So, the luminescence spectrum assumes a shape which depends on the linear coupling constant $S_f$ as in Eq. (15). At T ≠ 0, the coupling between the electronic transitions and the low frequency $\mathbf{n}_l$ modes induces changes in the luminescence spectrum like those reported for the absorption. In this case, the width of the emission profile as a function of the temperature can be expressed by Eq. (18), but $\langle \mathbf{n}_l \rangle$ indicates the mean frequency value of the $\mathbf{n}_l$ modes in the state (1). Moreover, as for the absorption, the conformational heterogeneity in amorphous materials causes a spread of the emission energies associated with the transition from (1) to (0), which results in the inhomogeneous broadening of the photoluminescence spectrum.

Generally, following light absorption, different excitation pathways can occur in a point defect. In Fig. 4, we depict a typical energetic level scheme consisting in a singlet ground state $S_0$, two singlet excited state $S_1$ and $S_2$ and the first triplet excited state $T_1$ [14]. The absorption $S_0 \to S_1$ and the fluorescence $S_1 \to S_0$ has been already discussed through the Fig. 3. At variance, if the system is excited to an electronic state $S_2$, it rapidly relaxes to the lowest vibrational levels of $S_1$. This process is called internal conversion and it occurs in

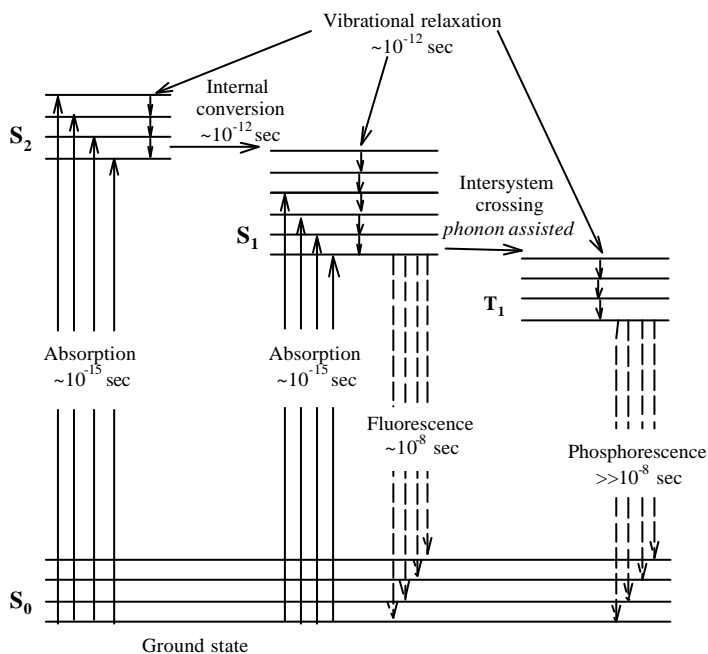

**Fig. 4** Excitation and relaxation pathways involving the ground and the first excited states localized on a point defect

~$10^{-12}$ sec. As the fluorescence lifetime is ~$10^{-8}$ sec, the internal conversion is complete before the radiative emission from $S_2$ to a lower singlet state so that the transition $S_0 \rightarrow S_2$ is able to excite the emission $S_1 \rightarrow S_0$. Finally, when the system is in the $S_1$ state, it can undergo a radiation-less transition to the first triplet state $T_1$ of lower energy. This non radiative conversion mechanism, known as intersystem crossing process, can be thermally activated by the interaction of luminescent defect with the lattice (phonon assisted process) [8]. The relaxation of $T_1$ towards the ground state causes a light emission (phosphorescence) at energies lower than the fluorescence. As the transition $T_1 \rightarrow S_0$ is forbidden, the phosphorescence lifetimes ($10^{-1}$-$10^{-5}$ sec) are several orders of magnitude longer than those of fluorescence. Therefore, the light absorption due to the transitions from the ground state $S_0$ to higher excited states $S_1$ and/or $S_2$ can excite at least two luminescence bands: one associated to the $S_1 \rightarrow S_0$ transition (fluorescence) and the other associated to the $T_1 \rightarrow S_0$ transition (phosphorescence).

## 3. Experimental method

### 3.1 Samples

In this work, we investigated a set of silica specimens of commercial origin, chosen so as to cover a wide spectrum of preparation techniques. These samples can be grouped in four standard silica types according to the early *Hetherington* classification [15]:

*Type I* natural dry silica is obtained by fusion of powder of quartz crystal via electric melting in vacuum or in an inert gas at low pressure. It contains negligible OH amount but about the same metallic impurities such as Al, Ge or alkali, totally of the order of 10 part per millions (ppm) by weight as the unfused raw material.

*Type II* natural wet is prepared by fusion of quartz crystal in a flame. This material has higher chemical purity than the type I because some impurities are volatilized in the flame but as it is prepared in a water-vapor atmosphere it contains nearly 150 ppm of OH groups.

*Type III* synthetic wet is made by the vapor-phase hydrolysis of pure silicon compounds such as $SiCl_4$. It contains the highest OH content (up to 1000 ppm) but it is virtually free from metallic impurities.

*Type IV* synthetic dry is obtained by the reaction of $O_2$ with $SiCl_4$ in a water-vapor-free plasma. In this way, the concentration of OH is reduced to less than 1 ppm but excess of Cl and oxygen in the form of -O-O- linkages are present.

Table 1. Sample list: sample name, silica type, and nominal OH content

| Sample Name | Type | OH (ppm) |
| --- | --- | --- |
| Infrasil 301 (I301) by Heraeus | Natural dry (I) | ≤ 8 |
| Puropsil QS (QPA) by Quartz & Silice | Natural dry (I) | 15 |
| Silica EQ906 by Quartz & Silice | Natural dry (I) | 20 |
| Silica EQ912 by Quartz & Silice | Natural dry (I) | 15 |
| Vitreosil (VTS) by TSL | Natural wet (II) | 150 |
| Homosil (HM) by Heraeus | Natural wet (II) | 150 |
| Herasil 1 (H1) by Heraeus | Natural wet (II) | 150 |
| Herasil 3 (H3) by Heraeus | Natural wet (II) | 150 |
| Suprasil 1 (S1) by Heraeus | Synthetic wet (III) | 1000 |
| Suprasil 311 (S311) by Heraeus | Synthetic wet (III) | 200 |
| Suprasil 300 (S300) by Heraeus | Synthetic dry (IV) | <1 |

In Table 1 we list the investigated silica types with their name and OH content. The materials Infrasil (I), Herasil (H), Homosil (HM), and Suprasil (S) were supplied by Heraeus [16]; silica EQ (EQ) were supplied by Quartz&Silice [17]; Vitreosil (VTS) was supplied by TSL [18]. All samples used in our measurements have sizes of 5×5×1 mm$^3$ with the major surfaces optically polished.

*3.2 Experimental techniques*

Absorption measurements at a wavelength between 190 to 340 nm, corresponding to 4.0-6.5 eV, were performed at room temperature using a JASCO V-570 double-beam spectrometer.

Photoluminescence emission (PL) and excitation (PLE) spectra, measured in steady state regime, were obtained by a Jasco PF-770 instrument, mounting a Xenon lamp of 150 W as light source. The samples were placed with the major faces at 45° with respect to the exciting beam and the PL light collected in the direction opposite to the reflected beam (45°-backscattering-geometry). The obtained excitation and emission spectra are corrected for wavelength dependent effects as spectral density of the source and the spectral response of detecting system.

PL and PLE measurements as a function of the temperature were also carried out by mounting the samples in a continuous flow helium cryostat (Oxford Optistat$^{CF}$), equipped with four optical windows and a temperature control (Oxford ITC503). The temperature could be varied from 4.2 up to 350 K and at the required temperature; the spectra were recorded after 10 min for thermal equilibrium.

Lifetime measurements were performed by using the synchrotron radiation (SR) at the SUPERLUMI station on the I-beamline of HASYLAB at DESY (Hamburg, Germany. The time decay of the transient PL was measured using 512 channels for scanning the time interval of 192 ns between adjacent SR pulses, pulse width 0.5 ns. These measurements were carried out at T=300 K and at T= 10 K.

**4. Results**

*4.1 PL activity in natural silica excited in the UV range: emissions at 3.1 eV and 4.2 eV*

In this section we review the experimental results concerning the optical absorption and the photoluminescence activity excited in the UV range in our natural silica [19-22] aiming to characterize the two PL bands centered at ~3.1 eV and ~4.2 eV and to make clear some aspects regarding their excitation mechanisms and their temperature dependence.

In Fig. 5 is reported the UV absorption spectrum detected in the I301 sample. The main absorption band ($B_{2\beta}$ band [23]) is characterized by a peak energy value $E_0$=5.15±0.01 eV, by a full width at half maximum FWHM=0.46±0.02 eV and by an amplitude at maximum $\alpha_{max}$=0.46±0.02 cm$^{-1}$, as obtained by a best fit procedure in *Gaussian* components. We note that the spectral components in the blue side of the spectrum are attributed to the tail of other OA bands located at higher energies and are taken into account in the fit procedure.

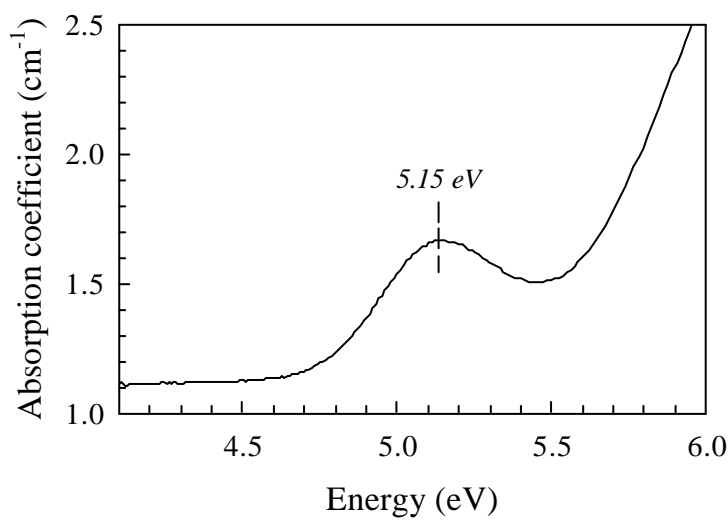

**Fig. 5** UV absorption profile for the natural silica I301. A band peaked at 5.15 eV ($B_{2\beta}$ band) is recognizable in the spectrum. Taken from *J. Non-Cryst. Solids* **245** (1999) 190-195, copyright 1999 by Elsevier Science.

The PL stationary activity excited at 5.0 eV at room temperature for the same sample is reported in Fig. 6. This spectrum exhibits two PL emissions: the first, at low energies, is centered at $E_0=3.16\pm0.01$ eV (with a FWHM=$0.43\pm0.02$ eV), the second is peaked at $E_0=4.26\pm0.01$ eV (with a FWHM=$0.48\pm0.02$ eV). These two emissions are labeled as the $\beta$ and $\alpha_E$ bands, respectively [24]. In the same figure, the PLE profiles for the emissions at 3.16 eV and 4.26 eV are also displayed. Apart from the small energy shift (0.05 eV), the excitation spectra of these two PL bands are quite similar to each other and have nearly the same FWHM (0.46 eV) as the $B_{2\beta}$ absorption band. We stress that the difference between the peak energies of the PLE spectra of $\beta$ and $\alpha_E$ can be related to the inhomogeneous distribution of the relaxation rates of the excited states [7, 22, 25, 26] in disordered systems such as the amorphous $SiO_2$.

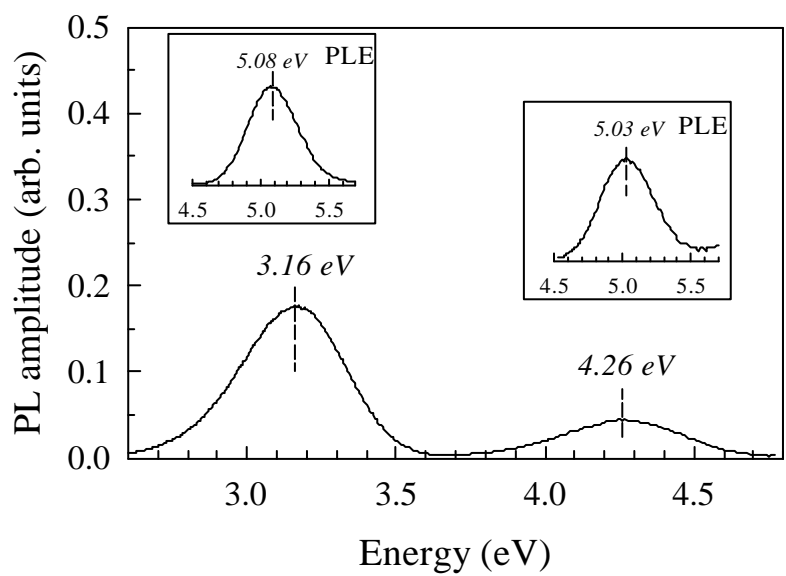

**Fig. 6** Photoluminescence spectrum excited at 5.0 eV for the I301 sample. Two main emissions are detected, centered at 3.16 eV ($\beta$) and 4.26 eV ($\alpha_E$), respectively. The insets show the corresponding excitation profiles of $\beta$ and $\alpha_E$.

**Table 2** Relevant spectral parameters (peak energy $E_0$, full width at half maximum FWHM and intensity I) of the absorption and luminescence bands as detected in natural silica samples. $E_0$ and FWHM are affected by an error of 0.01 and 0.02 eV, respectively, while $M_0$ is measured within an uncertainty of 5%.

| | OA band $B_{2\beta}$ | | | PL band $\beta$ | | | PL band $\alpha_E$ | | |
|---|---|---|---|---|---|---|---|---|---|
| Sample | $E_0$ (eV) | FWHM (eV) | I (cm$^{-1}$·eV) | $E_0$ (eV) | FWHM (eV) | I (a.u.) | $E_0$ (eV) | FWHM (eV) | I (a.u.) |
| I301 | 5.15 | 0.46 | 0.24 | 3.16 | 0.42 | 0.12 | 4.26 | 0.46 | 0.028 |
| EQ906 | 5.14 | 0.42 | 0.20 | 3.16 | 0.41 | 0.095 | 4.27 | 0.44 | 0.023 |
| EQ912 | 5.13 | 0.40 | 0.06 | 3.17 | 0.41 | 0.028 | 4.26 | 0.45 | 0.0059 |
| QPA | 5.13 | 0.41 | 0.07 | 3.16 | 0.41 | 0.036 | 4.27 | 0.44 | 0.0080 |
| VTS | 5.15 | 0.44 | 0.09 | 3.16 | 0.41 | 0.034 | 4.26 | 0.43 | 0.0077 |
| H1 | 5.16 | 0.46 | 0.21 | 3.16 | 0.42 | 0.10 | 4.27 | 0.44 | 0.024 |
| H3 | 5.12 | 0.43 | 0.08 | 3.16 | 0.41 | 0.035 | 4.26 | 0.44 | 0.0072 |
| HM | 5.15 | 0.47 | 0.19 | 3.16 | 0.41 | 0.079 | 4.27 | 0.44 | 0.018 |

The optical activity observed in the I301 sample, characterized by the OA band $B_{2\beta}$ and the two PL emissions $\beta$ and $\alpha_E$, is common to all natural silica samples. We stress that these optical bands are not detected in synthetic silica types [27, 28].

In table 2, we list the experimental values of the relevant quantities of these optical transitions. The integrated intensity $I$ or area of the bands is measured as the zero-th ($M_0$) energy moment of their spectral distribution $f(E)$ according to the definition: $I = M_0 = \int_{-\infty}^{\infty} f(E)dE$. It is worth to note that the spectral parameters (peak position $E_0$ and FWHM) of the bands $B_{2\beta}$, $\beta$ and $\alpha_E$ are the same in all investigated samples, within the experimental uncertainty. At variance, the integrated intensities depend on the specific sample, but they keep a strict correlation. This fact is evidenced in Fig. 7 (a) where the sum of the areas of the two PL bands $\beta$ and $\alpha_E$ is reported as a function of the $B_{2\beta}$ area, as measured in the eight silica samples considered here. Moreover, in Fig. 7 (b) the $\beta$ band area is plotted versus the $\alpha_E$ band one. The best linear fit is shown in both Figs. 7 (a) and (b) (line curves) and indicates that both the area ratio between the whole PL activity and the OA band $B_{2\beta}$ and between the two emissions $\beta$ and $\alpha_E$ areas is independent on the silica type. On the basis of these evidences, we can assert that the whole optical activity, here referred to as B type, originates from the same defect.

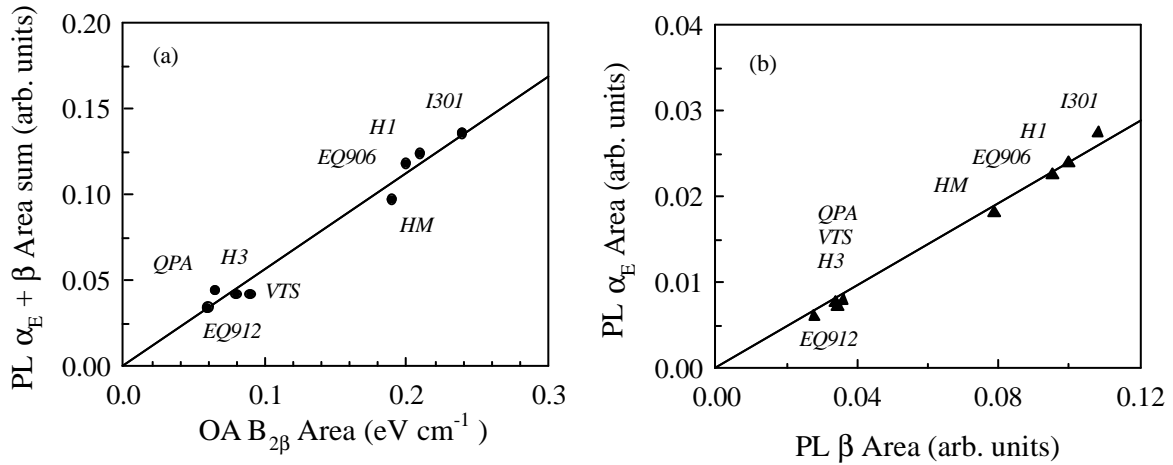

**Fig. 7** Correlation between the absorption and luminescence bands in our natural silica samples. The areas sum of the emissions $\alpha_E$ and $\beta$ versus the area of the absorption $B_{2\beta}$ (a). Intensity of the $\alpha_E$ band versus the intensity of the $\beta$ band. Taken from *Phys. Rev. B* **60** (1999) 11475-11481, copyright 1999 by the American Physical Society.

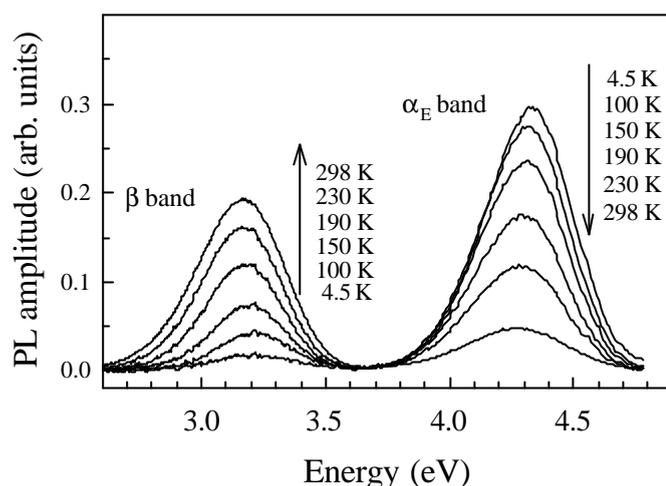

**Fig. 8** Emission spectra detected at various temperature in the sample I301 under excitation at 5.0 eV. Arrows indicate the changes induced on increasing the temperature. Taken from *Phys. Rev. B* **60** (1999) 11475-11481, copyright 1999 by the American Physical Society.

*4.2   Temperature dependence: intersystem crossing process*

The study of temperature effects on the PL activity is a useful tool to obtain information on the fine details of the processes involved in the excitation pathways. With this aim, we have investigated the temperature dependence of the $\beta$ and $\alpha_E$ emissions in the range 4.5-300 K. Fig. 8 shows the PL spectra excited at 5.0 eV, at various temperatures, for the I301 sample. The two PL bands exhibit an opposite behavior: on increasing the temperature, the $\alpha_E$ band decreases while the $\beta$ band increases starting from a near-to-zero value at low temperature.

The anticorrelated dependence on T of the $\alpha_E$ and $\beta$ emissions is summarized in Fig. 9. We observe that the sum of integrated intensities of the two PL bands is constant in the temperature range investigated.

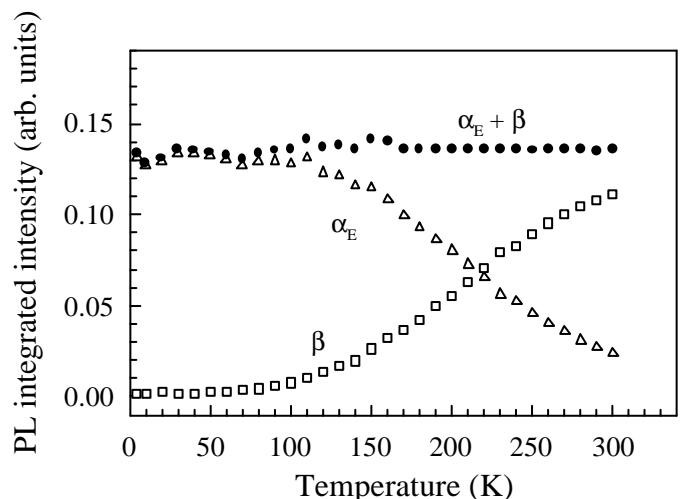

**Fig. 9** Integrated intensities of the $\alpha_E$ and $\beta$ PL bands as a function of the temperature. The sum of the two integrated intensities is also shown

The above reported data can be taken into account by the energy level scheme shown in Fig. 10 consisting in a singlet ground state $S_0$ and in two excited states of singlet $S_1$ and triplet $T_1$. For the sake of clarity, the radiative and non radiative processes with the relative rates are also indicated. We note that this quite general diagram was proposed by *L.N. Skuja* [25] to take into account the energetic levels of a defect consisting of a twofold coordinated Ge. According to the reported scheme, the $\alpha_E$ band is associated to the allowed $S_1 \rightarrow S_0$ fluorescence transition at a rate $K_r^F$, in agreement with its lifetime of the order of ns [21, 22]. On the other hand, the β emission is ascribed to the spin-rule forbidden $T_1 \rightarrow S_0$ phosphorescence transition at a rate $K_r^P$, in agreement with its longer decay time (~110 μs) [25]. Both emissions can be excited by the OA band $B_{2\beta}$ related to the $S_0 \rightarrow S_1$ transition: the $\alpha_E$ band directly and the β band via an intersystem crossing process between $S_1$ and $T_1$.

In this scheme, the intensities of the two emissions are proportional to the populations $N^{S_1}$ and $N^{T_1}$ of the states $S_1$ and $T_1$, and to the radiative rate $K_r^F$ and $K_r^P$, respectively, according to:

$$I^F \propto K_r^F \cdot N^{S_1} \tag{23}$$

$$I^P \propto K_r^P \cdot N^{T_1} \tag{24}$$

Moreover, the rate equations of $N^{S_1}$ and $N^{T_1}$ can be written as:

$$\frac{dN^{S_1}}{dt} = I_0 \cdot \left[1 - e^{-\alpha \cdot d}\right] - \left[K_r^F + K_{nr}^F + K_{ISC}\right] \cdot N^{S_1} \tag{25}$$

$$\frac{dN^{T_1}}{dt} = K_{ISC} \cdot N^{S_1} - \left[K_r^P + K_{nr}^P\right] \cdot N^{T_1} \tag{26}$$

where $I_0 \cdot \left[1 - e^{-\alpha \cdot d}\right]$ represents the intensity of the light absorbed by ground state defects.

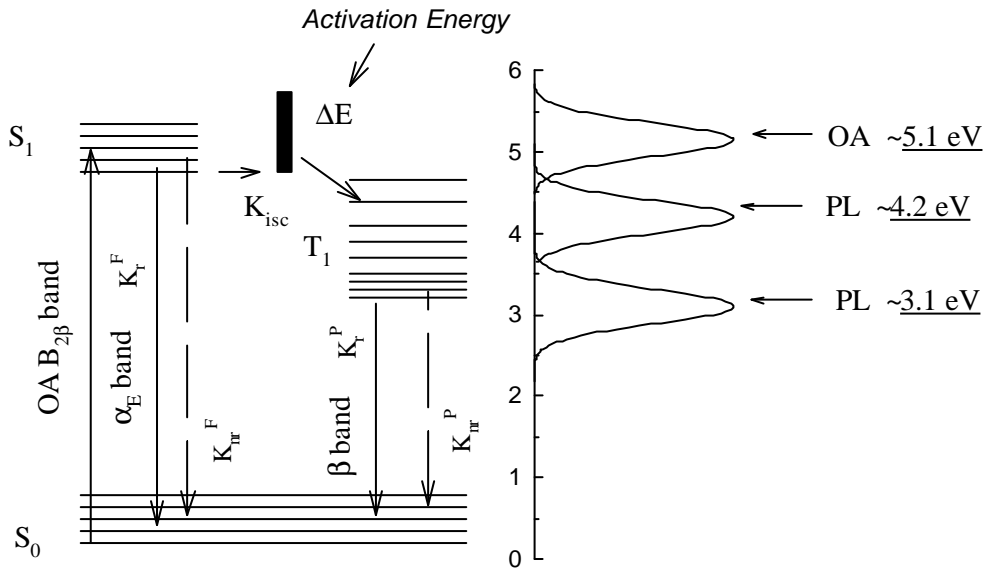

**Fig. 10** Outline of the electronic levels and of the related radiative and non radiative transitions accounting for the absorption at ~5.1 eV ($B_{2\beta}$) and the emissions at ~4.2 eV ($\alpha_E$) and ~3.1 eV (β).

In stationary (time-independent) condition, as obtained in our experiments under continuous light excitation, the first member in Eqs. (25) and (26) is reduced to zero and, combining with Eqs. (23) and (24), we get the steady-state (ss) solutions:

$$I_{ss}^F \propto K_r^F \cdot N_{ss}^{S_1} = Q^F \cdot I_0 \cdot \left[1 - e^{-\alpha l}\right] \quad (27)$$

$$I_{ss}^P \propto K_r^P \cdot N_{ss}^{T_1} = Q^P \cdot N_{ss}^{S_1} \cdot K_{ISC} \quad (28)$$

where $Q^F$ and $Q^P$ are the quantum yields of fluorescence and phosphorescence, respectively. They represent the fraction of chromophores that decay through radiative emission and are expressed by the ratio of the kinetic reaction constant of radiative decay and the sum of the kinetic reaction constant of all the decay processes:

$$Q^F = \frac{K_r^F}{K_r^F + K_{nr}^F + K_{ISC}} \quad (29)$$

$$Q^P = \frac{K_r^P}{K_r^P + K_{nr}^P} \quad (30)$$

Now, we wish to address to the thermal evolution of the two PL bands evidenced in Figs. 8 and 9. According to the energetic level diagram outlined in Fig. 10, the interconversion between the two excited states $S_1$ and $T_1$ is related to the intersystem crossing process. As reported in Section 2, the efficiency of this non-radiative mechanism is expected to depend upon the temperature as it arises from the interaction with lattice dynamics (phonon assisted process). In this way, changes induced by the temperature on $K_{ISC}$ could imply changes in the $S_1 \rightarrow S_0$ fluorescence quantum yield $Q^F$ and, according to Eqs. (27) and (28), opposite changes in the $T_1 \rightarrow S_0$ phosphorescence intensity.

We can suppose, as a first approximation, that the intersystem crossing process is governed by the presence of an activation barrier, so that the rate $K_{ISC}$ depends upon temperature according to the *Arrhenius* law [9]:

$$K_{ISC} = K_0 \cdot \exp(-\Delta E/K_B T) \quad (31)$$

where $\Delta E$ is the activation energy.

Moreover, since the sum of the integrated intensities of the two PL bands is constant in the temperature range 4.5-300 K (Fig. 9), we can assume that the temperature effects due to $K_{nr}^F$ and $K_{nr}^P$ are negligible in comparison with $K_{ISC}$. To corroborate this statement we also note that the lifetime of the β emission is temperature independent, as reported in [25], so evidencing that $K_{nr}^P << K_r^P$.

In this simplified scheme, the ratio $\eta$ between the integrated intensities (zero-th moment $M_0$) of $\alpha_E$ and β is expected to be:

$$\eta = \frac{[M_0]_{\alpha_E}}{[M_0]_{\beta}} \cong \frac{K_r^F}{K_{ISC}} = A_0 \cdot \exp\left(\frac{\Delta E}{K_B T}\right) \quad (32)$$

with $A_0 = K_r^F / K_0$.

To make this interpretation quantitative, we report in Fig. 11 the temperature dependence of $\eta$. The scales used in the graph make easier the comparison with the Eq. (32). As shown in the inset, a simple *Arrhenius* law is obeyed only at high temperature (T>150 K) with $\Delta E$=0.078±0.003 eV and $A_0$=(1.4±0.1)×10$^{-2}$. At variance, below 120 K, the ratio between the integrated intensities of $\alpha_E$ and β tends to a constant value and the intersystem crossing process becomes temperature independent.

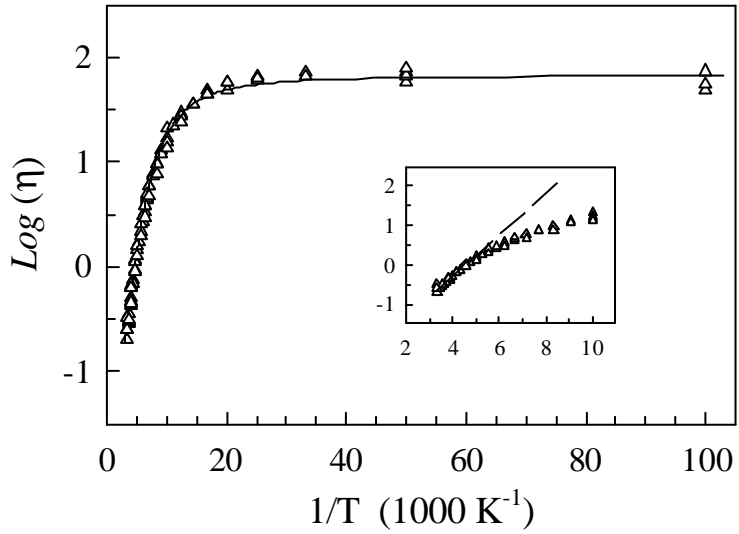

**Fig. 11** Temperature dependence of the ratio η between the integrated intensities of the $\alpha_E$ and β bands. The continuous line is a guide to the eye. The inset shows the initial part of the curve, in the temperature range 100 to 300 K. In this case, the dashed line results from a fit of Eq. (32) for T >150 K.

We stress that the knowledge of the $\eta$ behavior allows us to take into account the temperature dependence of the PL bands trough the ratio between $K_{ISC}$ and $K_r^F$. In particular, we get that the $K_{ISC}$ value is ~4·$K_r^F$ at T=300 K whereas it is ~0.02·$K_r^F$ for T<120 K. Then, in agreement with the data of Fig. 9, the $\alpha_E$ fluorescence quantum yield $Q^f$, expressed by Eq. (29) without the term $K_{nr}^F$, decreases by a factor ~5 on increasing the temperature from 4.5 to 300 K.

A final remark concerns the freezing of the intersystem crossing process at T<120 K. We observe that deviations from the *Arrhenius* law of $K_{ISC}$ are known to occur in other systems, like e.g. proteins [29, 30], and interpreted in terms of a distribution of the energy activation barriers $\Delta E$. In this frame, we can qualitatively interpret our results by hypothesizing that among the site-to-site non equivalent defects in natural silica, there is a small faction of B-active centers with $\Delta E$ values sufficient low to populate the $T_1$ state also at very low temperature.

*4.3 Lifetimes measurements*

A further experimental analysis of the luminescence activity of B-centers was performed by detecting the kinetic behavior of the $\alpha_E$ fluorescence under pulsed excitation. According to Eqs. (25) and (23), if the exciting light is abruptly switched off ($I_0$=0) we get:

$$\frac{dN^{S_1}}{dt} = -\left[K_r^F + K_{nr}^F(T) + K_{ISC}(T)\right] \cdot N^{S_1} \qquad (33)$$

and the PL signal decays by following the single exponential low:

$$I^F \propto K_r^F \cdot N^{S_1}(0) \cdot e^{-\frac{t}{\tau^F}} \qquad (34)$$

where $N^{S_1}(0)$ is the initial population in the excited state and $t^F = \left[K_r^F + K_{ISC}\right]^{-1}$ is the lifetime, assuming $K_{nr}^F \ll K_r^F$ and $K_{nr}^F \ll K_{ISC}$. Therefore, lifetime measurements can be useful to find out the temperature influence on the competition between radiative and the intersystem crossing rates arising from the $S_1$ state.

As shown in Fig. 12, the decay of $\alpha_E$ under excitation at 5.0 eV evidences relevant variation on changing the temperature. At T=10 K (a) the time decay is well described by a single exponential law with a lifetime $\tau=7.3\pm0.1$ ns. We note that at this temperature the relaxation of the excited state $S_1$ is essentially a radiative process and the lifetime tends to the value $t \approx \left(K_r^F\right)^{-1}$, so we can derive $K_r^F \sim 1.4\times10^8$ s$^{-1}$. On the other hand, the faster decay of curve (b), evidences the contribution of the intersystem crossing process to the $S_1$ relaxation and the lifetime is expected to decrease down to $t = \left(K_r^F + K_{ISC}\right)^{-1}$. We observe that at this temperature, the decay is not a single exponential. Then, it is intuitive to ascribe such a peculiar relaxation kinetic to the distribution of $K_{ISC}$ rates, which are effective at room temperature.

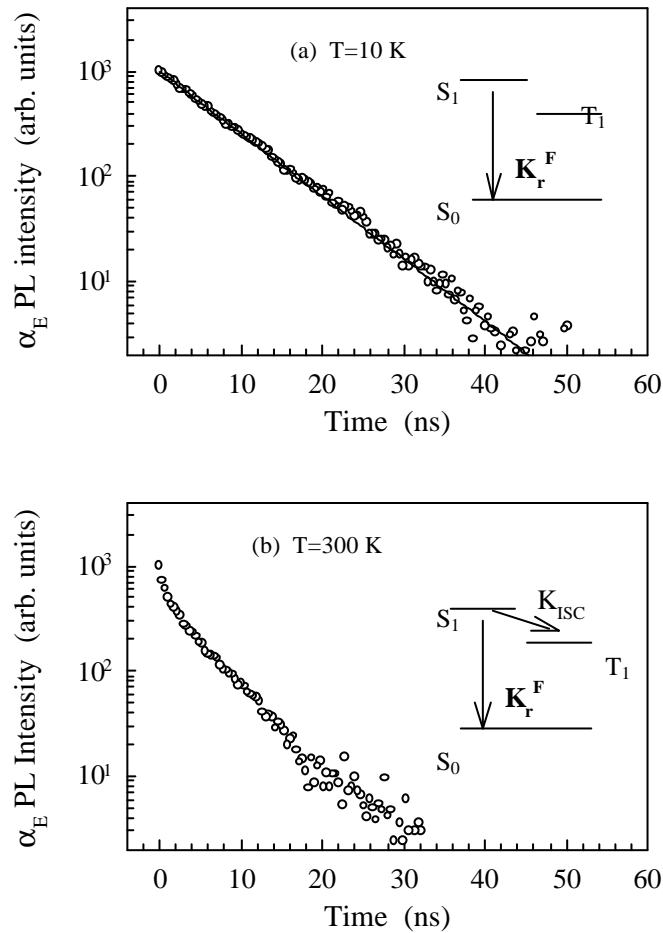

**Fig. 12** Time decay of the $\alpha_E$ emission at 4.2 eV observed under excitation at 5.0 eV at T=10 K (a) and T=300 K (b). Taken from *Phys. Rev. B* **60** (1999) 11475-11481, copyright 1999 by the American Physical Society.

*4.4 Temperature dependence of spectral moments: dynamic properties of the silica matrix*

To complete the study of the electronic properties of the optically B-active defects and investigate the influence of their environments, we have analyzed the energy moments of the spectral distributions, *f(E)*, relative to the $\alpha_E$, as a function of temperature [31]. Indeed, as outlined in section 2, the change of the optical band profile on varying the temperature allows to study the local dynamic properties and the heterogeneity of the matrix surrounding the chromophores [11, 12].

Fig. 13 shows the PL $\alpha_E$ band excited at 5.0 eV on varying the temperature, from room temperature to 4.5 K, in the I301 sample. In the paragraph *4.2* we have discussed about the marked temperature dependence of the $\alpha_E$ intensity (zero-th moment $M_0$), which is related to the effectiveness of the phonon, assisted intersystem crossing process. Here we focus our attention on the first ($M_1$) and second ($M_2$) spectral moments, which are calculated from the experimental data according to the following definitions:

$$M_1 = \frac{\int_{-\infty}^{\infty} E \cdot f(E) dE}{M_0} \qquad (35)$$

$$M_2 = \frac{\int_{-\infty}^{\infty} E^2 \cdot f(E) dE}{M_0} - M_1^2 \qquad (36)$$

We recall that $M_1$ is the mean value of the emission energy and measures the position of the PL band and $M_2$ is the mean square deviation of the spectral distribution *f(E)* and measures its width.

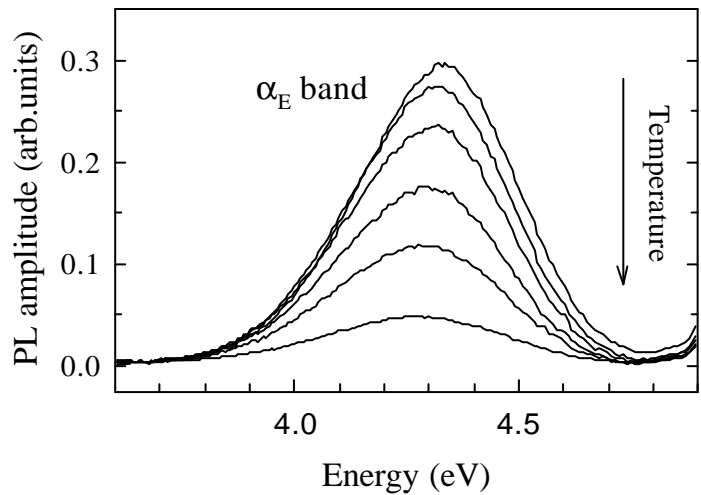

**Fig. 13** $\alpha_E$ band profile detected at various temperature. The arrow indicates the changes induced by increasing the temperature, curves refer to 4.5, 100, 150, 190, 230 and 298 K, respectively, from top downwards.

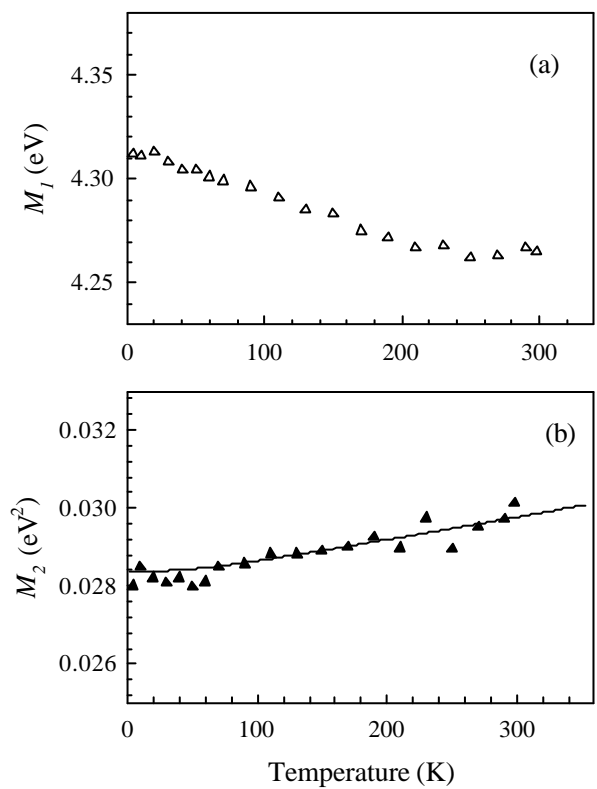

**Fig. 14** Thermal behavior of the $M_1$ (a) and $M_2$ (b) for the PL $\alpha_E$ band. The continuous line in (b) represents the fitting of Eq. (37). Taken from *J. Non Cryst. Solids* **232-234** (1998) 514-519, copyright 1998 by Elservier Sciences.

In Fig. 14, we report the thermal behavior of $M_1$ (a) and $M_2$ (b). Fig. (a) evidences the red-shift of the $\alpha_E$ emission on increasing the temperature, $M_1$ decreases from 4.32±0.01 eV at T=4.5 K to 4.26±0.01 eV at room temperature. We note that the temperature dependence of $M_1$ suggests the occurrence of a strong interaction of the electronic states with the surroundings. In this limit, the transition changes the vibrational frequencies of the ground and excited states (non linear coupling) resulting in temperature effects on the peak position of the optical bands [11, 12].

Fig. (b) shows the broadening of the PL band on increasing the temperature. At low temperature $M_2$ assumes the value of 0.028±0.001 (eV)$^2$, while it increases up to 0.031±0.001 (eV)$^2$ at room temperature. As already reported in section 2, we can take into account the thermal behavior of $M_2$, by assuming that the broadening of the PL emission of a single chromophore is dominated by its interaction with the local vibrational modes of the surrounding nuclei. The increase of the temperature causes an increase of the width of the band profile due to population of vibrational levels associated to the low frequency modes in their first excited states $S_1$. This effect adds to the coupling with the vibrational modes at high frequencies (temperature independent) whose energy distribution determines the shape of the PL bands. We stress that the broadening due to the electron-phonon coupling is a property of the individual center and it appears to be the same for any center, so we can call it homogeneous contribution. Moreover, in an amorphous material such as the silica, the site-to-site non-equivalence among the point defects embedded in the silica matrix results in an additional inhomogeneous broadening of the PL spectra. We note that the values of the kinetic decay constants, that are of the order of $2·10^8$ s$^{-1}$ (~$10^{-6}$ eV), allow us to exclude that the radiative decay rate contributes in a significant way to the observed bandwidth,

$\sqrt{M_2} \approx 0.17$ eV. Therefore, we can take into account the total broadening of the $\alpha_E$ emission band by the following equation:

$$M_2(T) = N_l \cdot S_l \cdot h^2 \langle \mathbf{n}_l \rangle^2 \coth\left(\frac{h\langle \mathbf{n}_l \rangle}{2K_B T}\right) + \mathbf{s}_{stat.}^2 \qquad (37)$$

where $\langle \mathbf{n} \rangle$, $N_l$ and $S_l$ are the mean frequency value, the number and the mean linear coupling constant of the bath of vibrational modes at low frequency (*Einstein* oscillator model), and the static width $\mathbf{s}_{stat.}$ accounts for the temperature independent broadening due to the *Poissonian* distribution of the high vibrational structures of a single center smeared by the broadening due to the inhomogeneous distribution of the transition energy of the luminescent centers. The continuous lines in Fig. 14 (b) represent the fit of Eq. (37) to the experimental data, the relative parameters being reported in Table 3.

**Table 3** Values of the parameters obtained by fitting the Eq. (37) to the thermal behavior of $M_2$ for the $\alpha_E$ band. Taken from *J. Non Cryst. Solids* **232-234** (1998) 514-519, copyright 1998 by Elservier Sciences

| $N_l \times S_l$ | $h\langle \mathbf{n} \rangle$ (meV) | $\mathbf{s}_{stat.}$ (meV) |
|---|---|---|
| 6±1 | 12±2 | 160±5 |

On the basis of these results, we can infer that, in the interaction with the surrounding silica matrix, the B-active defect has access to a low-frequency dynamics ($h\langle \mathbf{n} \rangle \sim 12$ meV). Moreover, we stress that the bandwidth of the $\alpha_E$ band is mainly related to $\mathbf{s}_{stat} \sim 160$ meV. In this respect, a direct trial to distinguish between the coupling with high frequency modes and the conformational heterogeneity of the PL centers can be achieved by using site-selective luminescence or spectral hole-burning techniques to suppress the inhomogeneous broadening effect [9]. These site-selective techniques have proven in detecting vibrational structures in non-bridging oxygen hole centers (NBOHC) in silica [32], whereas no such structures have been reported so far for other defects in silica.

## 5. Conclusions

In this paper, we have reviewed the optical properties of point defects in silica. In particular, our investigation has been focused on the study of PL emissions via their relationship with the absorption, their temperature dependence and their excitation pathways in the UV region. In this respect, the experimental evidences reported in section 4 contribute to outlining a clear picture of the optical features observed in a wide variety of natural silica samples and can be summarized as follows:

i) Two PL bands $\alpha_E$ at ~4.2 eV and $\beta$ at ~3.1 eV, related to the OA band at ~5.1 eV ($B_{2\beta}$) characterize the optical activity B associated to native defects present in natural silica.

ii) The temperature dependence of the PL intensities is governed by a phonon assisted intersystem crossing process occurring at the rate $K_{ISC}$ between the singlet and the triplet excited states from which the emissions $\alpha_E$ and $\beta$ originate, respectively.

iii) The departure from the simple *Arrhenius* law of the intersystem crossing thermal behavior and the no-single exponential time decay of $\alpha_E$ at room temperature indicate a distribution of $K_{ISC}$ rates, probably related to different environments of the centers in the amorphous structure of the silica.

iv) The analysis of the thermal behavior of the $\alpha_E$ emission profile (first and second spectral moment) enables to evaluate quantitatively the parameters of the electron-phonon coupling between the optically B active center and its environment.

As a final remark, the experiments and the related results described in this work prove that the parallel use of different luminescence techniques is a powerful tool to improve the understanding of the optical properties of defects in amorphous solids.

**Acknowledgments**

The author thanks S. Agnello, R. Boscaino, F.M. Gelardi and M. Leone for their collaboration and useful discussions during the experimental work and the preparation of this manuscript. This work is a part of a National Research Project supported by the Ministero della Ricerca Scientifica e Tecnologica, Roma, Italy.